\documentclass[twocolumn,aps,prc,superscriptaddress,reprint]{revtex4-1}
\usepackage{blindtext}
\usepackage{graphicx}
\usepackage{dcolumn}
\usepackage{bm}
\usepackage{booktabs}
\usepackage{color}
\usepackage{mathrsfs}
\usepackage{ulem}
\usepackage{verbatim}
\usepackage{indentfirst}
\usepackage{float}
\usepackage[colorlinks,
           bookmarks=true,
           linkcolor=blue,
           urlcolor=blue,
            anchorcolor=black,
            citecolor=blue
            ]{hyperref}
\usepackage[bf,raggedright,indentafter]{titlesec}
\usepackage{hypcap}

\usepackage{epsfig}

\begin{document}
\title{Cumulants of event-by-event net-strangeness distributions in Au+Au collisions at $\sqrt{s_\mathrm{NN}}$=7.7-200 GeV from UrQMD model }
\author{Chang Zhou}
\affiliation{ Key Laboratory of Quark\&Lepton Physics (MOE) and Institute of Particle Physics,\\
Central China Normal University, Wuhan 430079, China}
\author{Ji Xu}
\affiliation{ Key Laboratory of Quark\&Lepton Physics (MOE) and Institute of Particle Physics,\\
Central China Normal University, Wuhan 430079, China}
\author{Xiaofeng Luo}
\email{xfluo@mail.ccnu.edu.cn}
\affiliation{ Key Laboratory of Quark\&Lepton Physics (MOE) and Institute of Particle Physics,\\
Central China Normal University, Wuhan 430079, China}
\affiliation{Department of Physics and Astronomy, University of California, Los Angeles, California 90095, USA}
\author{Feng Liu}
\affiliation{ Key Laboratory of Quark\&Lepton Physics (MOE) and Institute of Particle Physics,\\
Central China Normal University, Wuhan 430079, China}
\begin{abstract}
Fluctuations of conserved quantities, such as baryon, electric charge and strangeness number, are sensitive observables in heavy-ion collisions to search for the QCD phase transition and critical point. In this paper, we performed a systematical analysis on the various cumulants and cumulant ratios of event-by-event net-strangeness distributions in Au+Au collisions at $\sqrt{s_{NN}}$=7.7, 11.5, 19.6, 27, 39, 62.4 and 200 GeV from UrQMD model. We performed a systematical study on the contributions from various strange baryons and mesons to the net-strangeness fluctuations. The results demonstrate that the cumulants and cumulant ratios of net-strangeness distributions extracted from different strange particles show very different centrality and energy dependence behavior. By comparing with the net-kaon fluctuations, we found that the strange baryons play an important role in the fluctuations of net-strangeness. This study can provide useful baselines to study the QCD phase transition and search for the QCD critical point by using the fluctuations of net-strangeness in heavy-ion collisions experiment. It can help us to understand non-critical physics contributions to the fluctuations of net-strangeness.
\end{abstract}
\maketitle

\section{Introduction}
One of the main goals of the high energy nuclear collisions is to explore the phase structure of strongly interacting hot and dense nuclear matter and map the quantum chromodynamics (QCD) phase diagram which can be displayed by the temperature ($T$) and baryon chemical potential ($\mu_{B}$). 
Finite temperature lattice quantum chromodynamics (LQCD) calculations at zero baryon chemical potential region predicted that the transition from the hadronic phase to quark-gluon plasma phase is a smooth crossover, ~\cite{Aoki:2006we,Aoki:2006br}, While at large $\mu_{B}$ and low temperature region, the finite density phase transition is of first order~\cite{Ejiri:2008xt,Endrodi:2011gv,deForcrand:2002hgr}. So, there should be an end point at the end of the first order phase transition boundary towards the crossover region~\cite{Stephanov:2004wx,Fodor:2004nz}. 

Fluctuations of conserved quantities, such as net-baryon (B), net-charge (Q) and net-strangeness (S),  have been predicted to be sensitive to the QCD phase transition and QCD critical point. Experimentally, one can measure various order moments (Variance($\sigma^2$), Skewness($S$), Kurtosis($\kappa$)) of the event-by-event conserved quantities distributions in heavy-ion collisions. These moments are sensitive to the correlation length ($\xi$) of the hot dense matter created in the heavy-ion collisions~\cite{Stephanov:2008qz,Athanasiou:2010kw,Hatta:2003wn} and also connected to the thermodynamic susceptibilities computed in Lattice QCD~\cite{Gavai:2010zn,Gavai:2008zr,Cheng:2008zh,Bazavov:2012vg,Ding:2015ona,Bazavov:2012jq,Friman:2011pf,Mukherjee:2015swa,Morita:2014fda,Alba:2017mqu,Noronha-Hostler:2016sje} and in the Hadron Resonance Gas (HRG)~\cite{Karsch:2010ck,Garg:2013ata,Fu:2013gga,Nahrgang:2014fza,Alba:2014eba,Alba:2015iva,Alba:2015lxa} model. 
These have been studied widely in experiment and theoretically~\cite{Luo:2009sx,Nahrgang:2015tva,Chen:2014ufa,Jiang:2015cnt,Gupta:2011wh,Asakawa:2009aj,Stephanov:2011pb,Thader:2016gpa}. Experimentally, strange hadrons in the final state production can provide deep insight into the characteristics of the system since they are not inherent inside the nuclei of the incoming beam. Thus, the yield ratios and fluctuations of strange particles have been studied at different experiments~\cite{Al2001An,Antinori:2004ee,Alt:2004kq,Adler:2002uv}. Experimentally, the STAR experiment has reported the cumulants of net-kaon (proxy for net-strangeness) multiplicity distributions at $\sqrt{s_\mathrm{NN}}$=7.7, 11.5, 14.5, 19.6, 27, 39, 62.4 and 200 GeV~\cite{Xu:2016hxf,Thader:2016gpa}.  However, the net-kaon is not a conserved quantity in QCD. We want to know, to what extend, the net-kaon fluctuations can be used as an approximation of fluctuations of net-strangeness in heavy-ion collisions. Thus, we calculated the cumulants of net-strangeness distributions in Au+Au collisions at RHIC BES energies by including different strange baryons and mesons with UrQMD model in version 2.3~\cite{Bleicher:1999xi}. This is to study the contribution from the strange baryons and mesons to the fluctuations of net-strangeness. This study can provide baselines and qualitative background estimates for the search for QCD phase transition and QCD critical point in relativistic heavy-ion collisions. 

This paper is organized as follows. In section II, we will introduce the UrQMD model. Then, we show the definition of cumulants and cumulant ratios in heavy-ion collisions in the section III. Furthermore, we present the net-strangeness fluctuation with the contributions from different strange particles in Au+Au collisions from the UrQMD calculations and discuss physical implications of these results in the section IV. Finally, the summary will be given in section V. 

\section{UrQMD Model}
The Ultrarelativistic Quantum Molecular Dynamics (UrQMD)~\cite{Bleicher:1999xi} approach is one of the microscopic transport models to describe subsequent individual hadron-hadron interactions and system evolution. Based on the covariant propagation of all hadrons with stochastic binary scattering, color string formation and resonance decay~\cite{Bleicher:1999xi}, UrQMD model can provides phase space  descriptions~\cite{Bass:1998ca} of different reaction mechanisms. At higher energies,e.g. $\sqrt{s_{NN}}> 5$ GeV, the quark and gluon degrees of freedom can not be neglected. And the excitation of color strings and their subsequent fragmentation into hadrons are the dominate mechanisms for the multiple production of particles. 

In addition, UrQMD approach can simulate hadron-hadron interactions at heavy-ion collisions with the entire available range of energies from SIS energy ($\sqrt{s_{NN}} = 2$ GeV) to RHIC top energy ($\sqrt{s_{NN}} = 200$ GeV) and the collision term in UrQMD model covers more than fifty baryon species and 45 meson species as well as their anti-particles~\cite{Bleicher:1999xi}.  The comparison of the data (this paper deals with net-strangeness fluctuations) onto those obtained from UrQMD model will tell about the contribution from the hadronic phase and its associated processes.

\section{Observables}
Experimentally, one can measure particle multiplicity in an event-by-event basis.  By measuring the final state strange particle and anti-particles in heavy-ion collisions, we can count the strange quark ($N_{s}$) and anti-strange quark number ($N_{\bar{s}}$) in those strange hadrons, respectively.  Different strange particles have different number of (anti-)strange quarks, e.g., the 
strange baryon $\Lambda$,  $\Xi $ and $\Omega$ consist of 1, 2 and 3 strange quarks, respectively, and the strange quark and anti-strange quark carry negative and positive strangeness quantum number, respectively. We use $N=N_{\bar{s}} - N_{s}$ to denote the number of net-strangeness in one event and $<N>=<N_{\bar{s}}> - <N_{s}>$ to denote the mean value of the net-strangeness over the whole sample, where $N_s$ and $N_{\bar s}$ represent the number of strange quark and anti-strange quark in one event ($\begin{array}{*{20}{c}}{{N_f} = \sum\limits_i^{} {n_i^f{p_i},} }&{f = \bar s,s} \end{array}$) and the $n_i^f$ are the strange ($f=s$) or anti-strange quark number ($f=\bar s$) for the strange particle $p_i$ in one event.

Then the deviation of $N$ from its mean value can be defined as $\delta N = N - <N>$. The various order cumulants of event-by-event distributions of the variable $N$ can be defined as follows,

\begin{eqnarray}
C_{1,N} = <N>\text{,}
\end{eqnarray}  
\begin{eqnarray}
C_{2,N} = <(\delta N)^{2}>\text{,}
\end{eqnarray}  
\begin{eqnarray}
C_{3,N} = <(\delta N)^{3}>\text{,}
\end{eqnarray}  
\begin{eqnarray}
C_{4,N} = <(\delta N)^{4}> - 3<(\delta N)^{2}>^{2}\text{.}
\end{eqnarray}  
Once we have the definition of cumulants, various moments of net-strangeness distribution can be written as,
\begin{eqnarray}
M = C_{1,N}\text{,}
\end{eqnarray}  
\begin{eqnarray}
\sigma^{2} = C_{2,N}\text{,}
\end{eqnarray}  
\begin{eqnarray}
S = \frac{C_{3,N}}{(C_{2,N})^{3/2}}=\frac{<(\delta N)^{3}>}{\sigma^{3}}\text{,}
\end{eqnarray}  
\begin{eqnarray}
\kappa = \frac{C_{4,N}}{(C_{2,N})^{2}}=\frac{<(\delta N)^{4}>}{\sigma^{4}}-3 \text{.}
\end{eqnarray}  
Statistically~\cite{Hald2000The}, various cumulants are used to describe the shape of a probability distribution. 
For instance, the variance ($\sigma^{2}$) characterizes the width of a distribution, while the skewness ($S$) and kurtosis ($\kappa$) are used to describe the asymmetry and peakness of a distribution, respectively.  Theoretical and QCD based model calculations show that the high order cumulants of conserved quantities, such as baryon, strangeness and electric charge number, are proportional to the high power of correlation length ($\xi$)~\cite{Athanasiou:2010kw,Gavai:2010zn}. 
\begin{eqnarray}
<(\delta N)^{2}> \sim \xi^{2}\text{,}
\end{eqnarray}  
\begin{eqnarray}
<(\delta N)^{3}> \sim \xi^{4.5}\text{,}
\end{eqnarray} 
\begin{eqnarray}
<(\delta N)^{4}>-3<(\delta N)^{2}>^{2} \sim \xi^{7}\text{.}
\end{eqnarray} 


Lattice QCD calculation tell us that the cumulants of conserved quantities are sensitive to the susceptibilities of the system~\cite{Ding:2015ona,Luo:2017faz},
\begin{eqnarray}
C_{n,N}=VT^{3}\chi^{(n)}_{N}(T,\mu_{N})\text{,}
\end{eqnarray} 
where $V$ is the volume of the system. Experimentally, it is very difficult to measure the volume of the collision system, so the cumulant ratios are constructed to remove the effect of system volume. 
The moment product $\kappa\sigma^{2}$ and $S\sigma$ can be expressed in terms of cumulant ratios: 
\begin{eqnarray}
\frac{\chi^{(3)}_{N}}{\chi^{(2)}_{N}}=\frac{C_{3,N}}{C_{2,N}}=(S\sigma)_{N}\text{,}
\end{eqnarray}
\begin{eqnarray}
\frac{\chi^{(4)}_{N}}{\chi^{(2)}_{N}}=\frac{C_{4,N}}{C_{2,N}}=(\kappa\sigma^{2})_{N}\text{.}\label{ks}
\end{eqnarray}

With above definitions, we can calculate various cumulants and cumulant ratios for the measured event-by-event net-particles multiplicity distributions.

\begin{figure*}[htb] 
  \begin{center}
    \includegraphics[height=8cm]{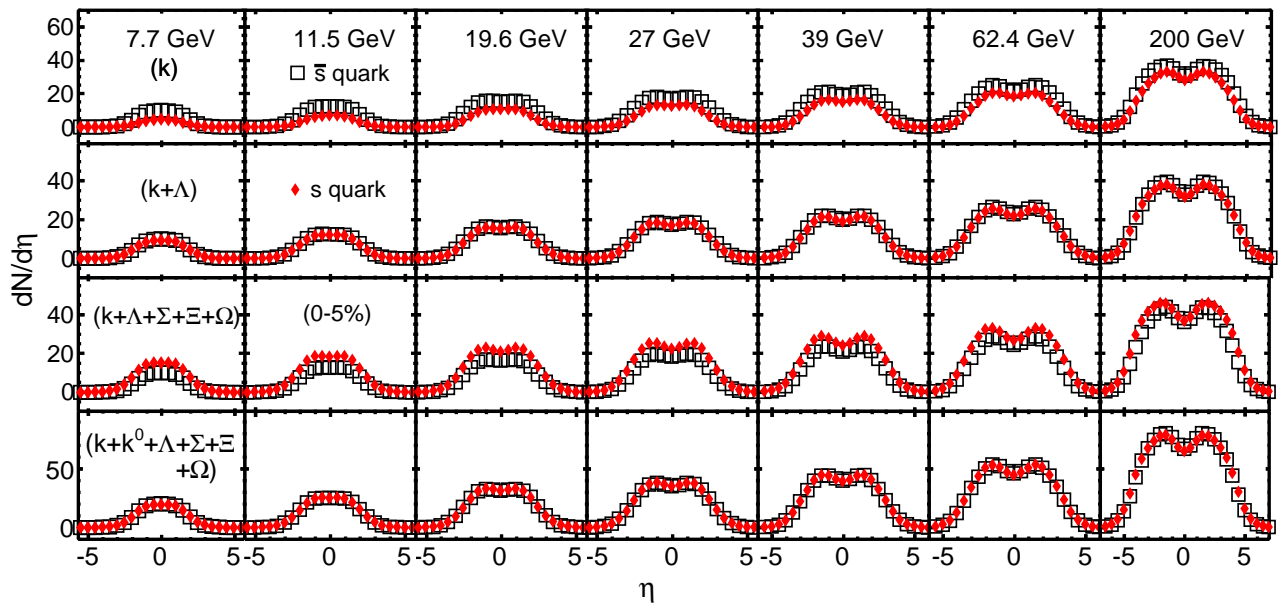}
    \caption{(color online) The $dN/d\eta$ distribution of strange and anti-strange quark multiplicities in four different cases in 0-5\% most central Au+Au collisions at $\sqrt{s_{NN}}$=7.7, 11.5, 19.6, 27, 39, 62.4 and 200 GeV from UrQMD model.}
   \label{fig:1}
  \end{center}
\end{figure*}

\begin{figure}[htb] 
\hspace{-1cm}
 \includegraphics[height=6.5cm]{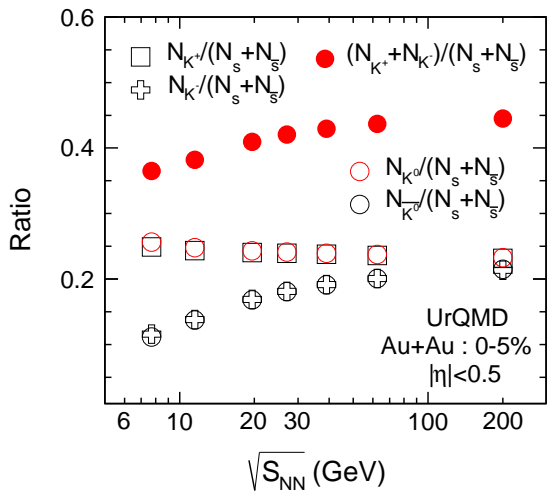}
\caption{(color online) Energy dependence of yields ratio for the most central (0-5\%) Au+Au collisions at mid-rapidity ($|\eta|<0.5$) from UrQMD model. }
\label{fig:2}
\end{figure}

\section{Results}\label{sec:3}

In this section, we present the centrality, rapidity and collision energy dependence of various cumulants ($C_{1}, C_{2}, C_{3}$ and $C_{4}$) and cumulant ratios ($\kappa\sigma^{2}, S\sigma$) of net-strangeness distributions for Au+Au collisions at $\sqrt{s_{NN}}$=7.7, 11.5, 19.6, 27, 39, 62.4 and 200 GeV from UrQMD model.
From low to high energies, the corresponding statistics are 35, 113, 113, 83, 135, 135 and 56 million minimum bias events, respectively. 

The statistical errors are estimated based on the Delta theorem~\cite{Luo:2011tp,Luo:2014rea}. To avoid auto-correlation, the collision centralities are determined by the (anti-)proton and charged pion multiplicities within pseudo-rapidity $|\eta|<1$. We perform our calculation with four cases ($(1) ~K, (2) ~K+\Lambda, (3) ~K+\Lambda+\Sigma+\Xi+\Omega, (4)~K+K^{0}+\Lambda+\Sigma+\Xi+\Omega $), where both the particle and anti-particles are included. For each case, we can calculate the cumulants of net-strangeness distributions. 
\begin{figure*}[htb] 
  \begin{center}
    \includegraphics[height=8cm]{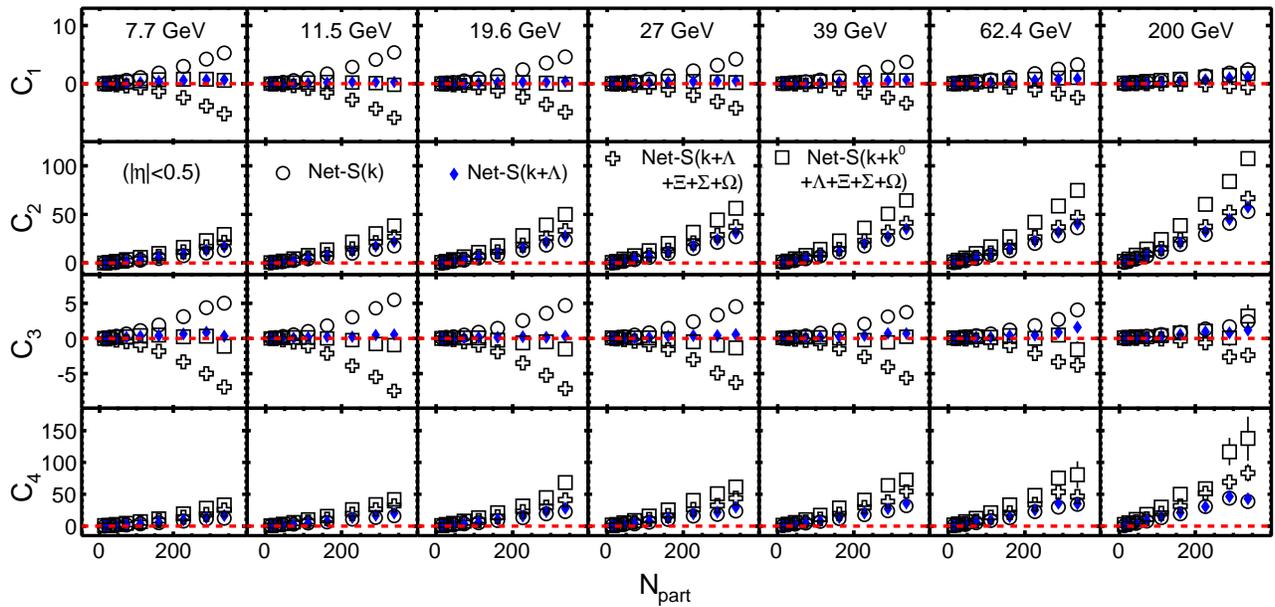}
    \caption{(color online) Various cumulants of net-strangeness multiplicity distributions as a function of $N_{part}$ at mid-rapidity region ($|\eta|<0.5$) for Au+Au collisions at $\sqrt{s_{NN}}$=7.7, 11.5, 19.6, 27, 39, 62.4 and 200 GeV with UrQMD model.} 
   \label{fig:3}
  \end{center}
\end{figure*}

Figure~\ref{fig:1} shows the pseudo-rapidity distributions ($dN/d\eta$) of strange quark and anti-strange quark for the most central (0-5\%) Au+Au collisions at $\sqrt{s_{NN}}$= 7.7 to 200 GeV calculated from the UrQMD model for the above four cases. The ${\left. {\frac{{dN}}{{d\eta }}} \right|_{\eta  = 0}}$ of strange and anti-strange quarks monotonously increase with increasing collision energy from 7.7 to 200 GeV for all the four cases. If one considers only the $K^{+}$ and $K^{-}$ (top row in Fig. ~\ref{fig:1} ), the $dN/d\eta$ distributions of the anti-strange quarks are above the strange quarks at all energies. The differences of $dN/d\eta$ between strange quark and anti-strange quark become smaller at higher energies. If we include the strange baryons, such as the case of ($K+\Lambda+\Sigma+\Xi+\Omega$), the $dN/d\eta$ distributions of strange quarks are slightly above the anti-strange quarks. 
This can be explained by the interplay between the associate production and pair production of $K^{+}$ and $K^{-}$ from lower to higher energies. 
At lower energies, the associate production from the reaction channel $NN\rightarrow N \Lambda K^{+}$ dominates the production of $K^{+}$ which leads to the number of sbar quarks being larger than the number of $s$ quarks. 
However, the $K^{+}$ and $K^{-}$  are mainly produced from pair production at higher energies, which means the number of $\bar{s}$ quark and $s$ quark are similar.

If we want to know to what extend the net-kaon fluctuations can reflect the fluctuations of net-strangeness,  the first step is to demonstrate the fraction of strangeness carried by $K^{+}$ and $K^{-}$ over the total strangeness. Figure~\ref{fig:2} shows the energy dependence of ratios, which are the number of total strangeness carried by kaons ($K^{+}$ and $K^{-}$) divided by the total strangeness from all strange particles at mid-rapidity in 0-5\% most central Au+Au collisions from UrQMD calculations. 
We found that the ratios of $N_{K^{-}}/N_{s+\bar{s}}$ and $N_{K^{+}+K^{-}}/N_{s+\bar{s}}$ have a smooth increase with  energy increasing from 7.7 to 200 GeV and the value of $N_{K^{+}+K^{-}}/N_{s+\bar{s}}$ at $\sqrt{s_\mathrm{NN}}$=200 GeV is about $45\%$. On the other hand, the ratios of $N_{K^{+}}/N_{s+\bar{s}}$ smoothly decrease with increasing energy. At low energies, such as 7.7, 11.5 and 19.6 GeV, the values of $N_{K^{+}}/N_{s+\bar{s}}$ are much larger than those of  $N_{K^{-}}/N_{s+\bar{s}}$ whereas the values of $N_{K^{+}}/N_{s+\bar{s}}$ and $N_{K^{-}}/N_{s+\bar{s}}$ are very close to each other at higher energies. The effects of the energy dependence of changing kaon production can also be explained by the kaon production mechanism. We also show the fraction of strangeness carried by $K^{0}$ and ${\bar K^0}$ over the total strangeness, which are similar to the charged kaons. This can be understood 
by the isospin balance between $u$ and $d$ quarks in the mid-rapidity of heavy-ion collisions. The yields between $K^{+}$ and $K^{0}$, $K^{-}$ and ${\bar K^0}$ should be very close to each other, respectively. 

Figure~\ref{fig:3} shows the centrality dependence of various cumulants of net-strangeness multiplicity distributions at mid-rapidity in Au+Au collisions at $\sqrt{s_{NN}}$=7.7 to 200 GeV from UrQMD calculations. Based on the similarity of the trends, those cumulants ($C_{1},C_{2},C_{3}$ and $C_{4}$) can be separated into odd order ($C_{1},C_{3}$) and even order cumulants ($C_{2},C_{4}$). The $C_{2}$ and $C_{4}$ show monotonically increase from peripheral collision to central collision and the even order cumulants of net-strangeness extracted from $K$ and $K+\Lambda$ have very close values. It is observed that $C_{1}$ and $C_{3}$ also have similar trend and the values of net-strangeness from $K+\Lambda$ and $K+K^{0}+\Lambda+\Xi+\Sigma+\Omega$ are close to zero. The net-strangeness number at initial state is zero, due to the strangeness conservation, the net-strangeness number should be also zero at finial state. The results indicate a better approximation for the real net-strangeness is reached by including more strange particles into the calculations. On the other hand, the odd order cumulants of net-strangeness from $K+\Lambda+\Xi+\Sigma+\Omega$ are negative. It is because that it has more number of strange baryons (like $\Lambda$, $\Xi$, $\Sigma$ and $\Omega$) than the number of anti-strange baryons especially at low energies. This explains why the odd order cumulants of net-strangeness ($N_{\bar{s}}-N_{s}$) remain negative.

\begin{figure*}[htb] 
  \begin{center}
    \includegraphics[height=8cm]{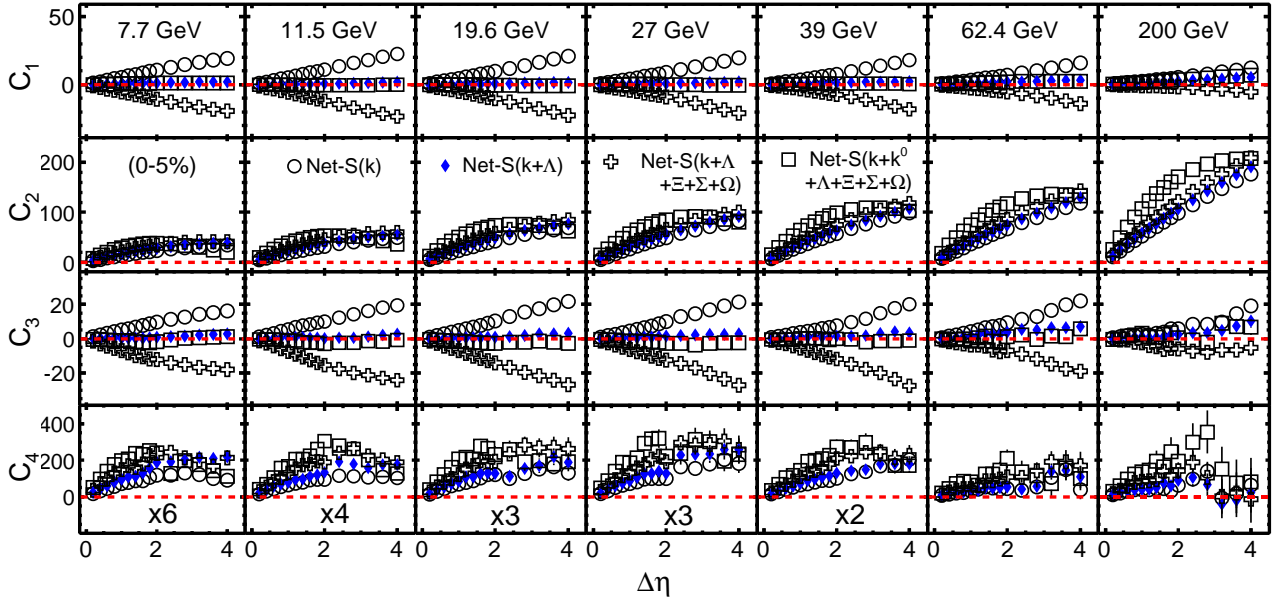}
    \caption{(color online) Pseudo-rapidity window size dependence of various cumulants of net-strangeness distributions for the most central (0-5\%) Au+Au collisions from $\sqrt{s_{NN}}$=7.7 to 200 GeV with UrQMD model.} 
   \label{fig:4}
  \end{center}
\end{figure*}

\begin{figure}[htb] 
\hspace{-1cm}
\includegraphics[height=7.0cm]{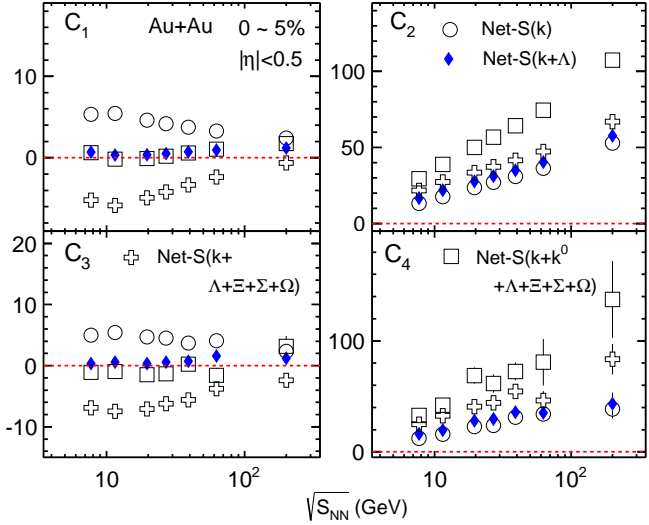}
\caption{(color online) Energy dependence of various cumulants of Net-strangeness multiplicity distributions at mid-rapidity region ($|\eta|<0.5$) for most central (0-5\%) Au+Au collisions with UrQMD model.} 
\label{fig:5}
\end{figure} 

Figure~\ref{fig:4} shows various cumulants of net-strangeness multiplicity distributions as a function of pseudo-rapidity window size for the 0-5\% most central Au+Au collisions at $\sqrt{s_\mathrm{NN}}$=7.7 to 200 GeV from UrQMD calculations. 
It is similar to the centrality dependence of various cumulants as is shown in Fig~\ref{fig:3}.  The odd order cumulants $C_{1}$ and $C_{3}$ show linear variation with the window size and the results from  $K+\Lambda+\Xi+\Sigma+\Omega$ remain negative due to the large number of strange quarks. For the even order cumulants, they show linear increase with increasing the rapidity window size. When $\Delta \eta$ is around 3, the even order cumulants can reach saturation and suppression, which can be understood by the effects of net-strangeness number conservation.
\begin{figure*}[htb] 
	\begin{center}
		\includegraphics[height=7cm]{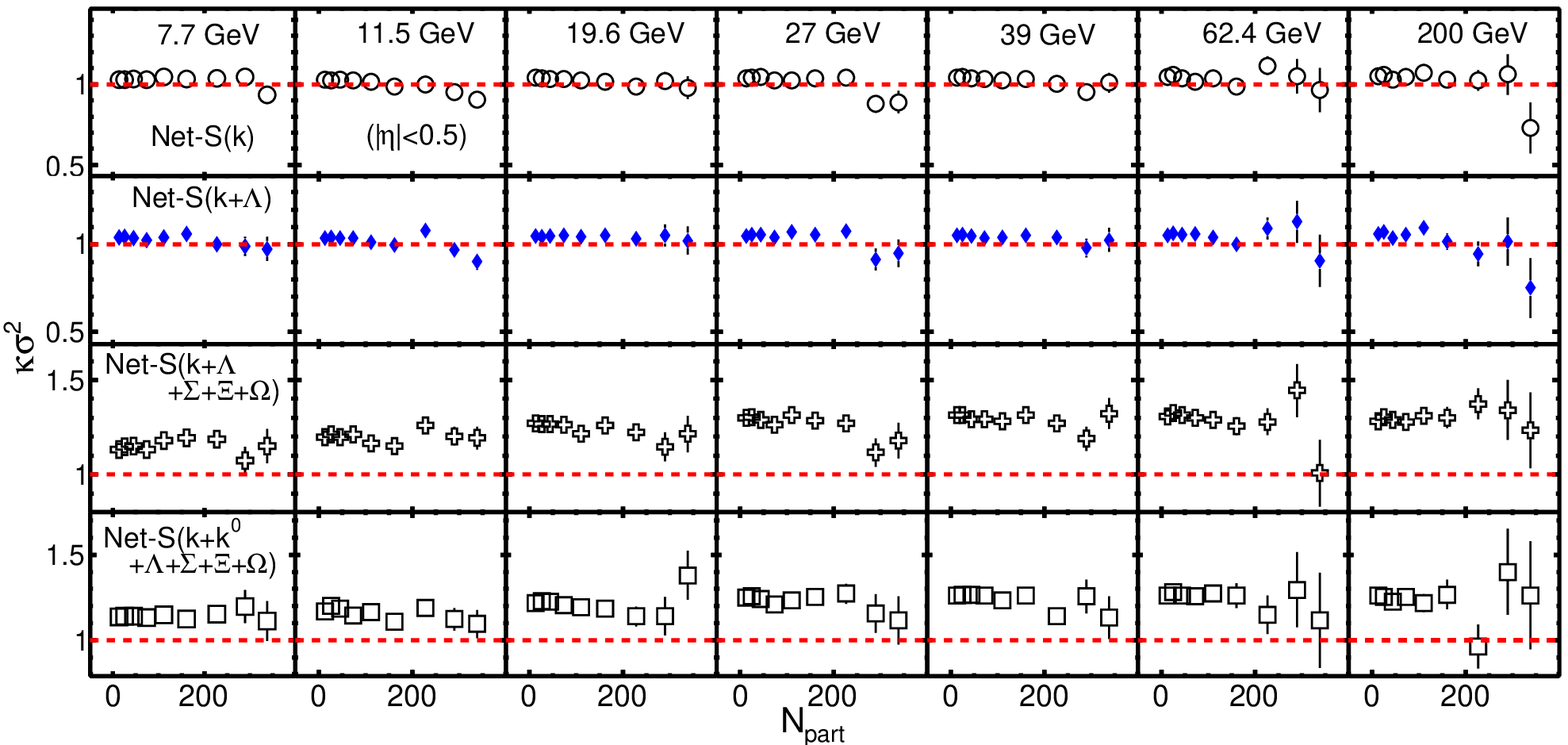}
		\caption{(color online) Centrality dependence of $\kappa\sigma^{2}$ in Au+Au collisions at $\sqrt{s_{NN}}$=7.7 to 200 GeV from UrQMD model.} 
		\label{fig:6}
	\end{center}
\end{figure*} 

\begin{figure*}[htb] 
	\begin{center}
		\includegraphics[height=7cm]{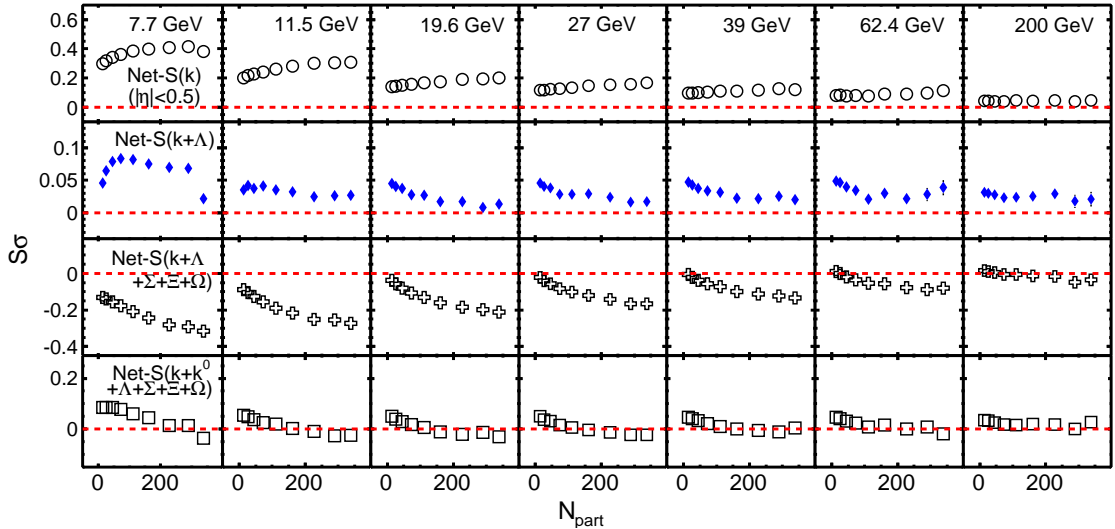}
		\caption{(color online) Centrality dependence of $S\sigma$ in the Au+Au collisions at $\sqrt{s_{NN}}$=7.7 to 200 GeV from UrQMD model.} 
		\label{fig:7}
	\end{center}
\end{figure*} 

\begin{figure*}[!htb] 
  \begin{center}
    \includegraphics[height=6.5cm]{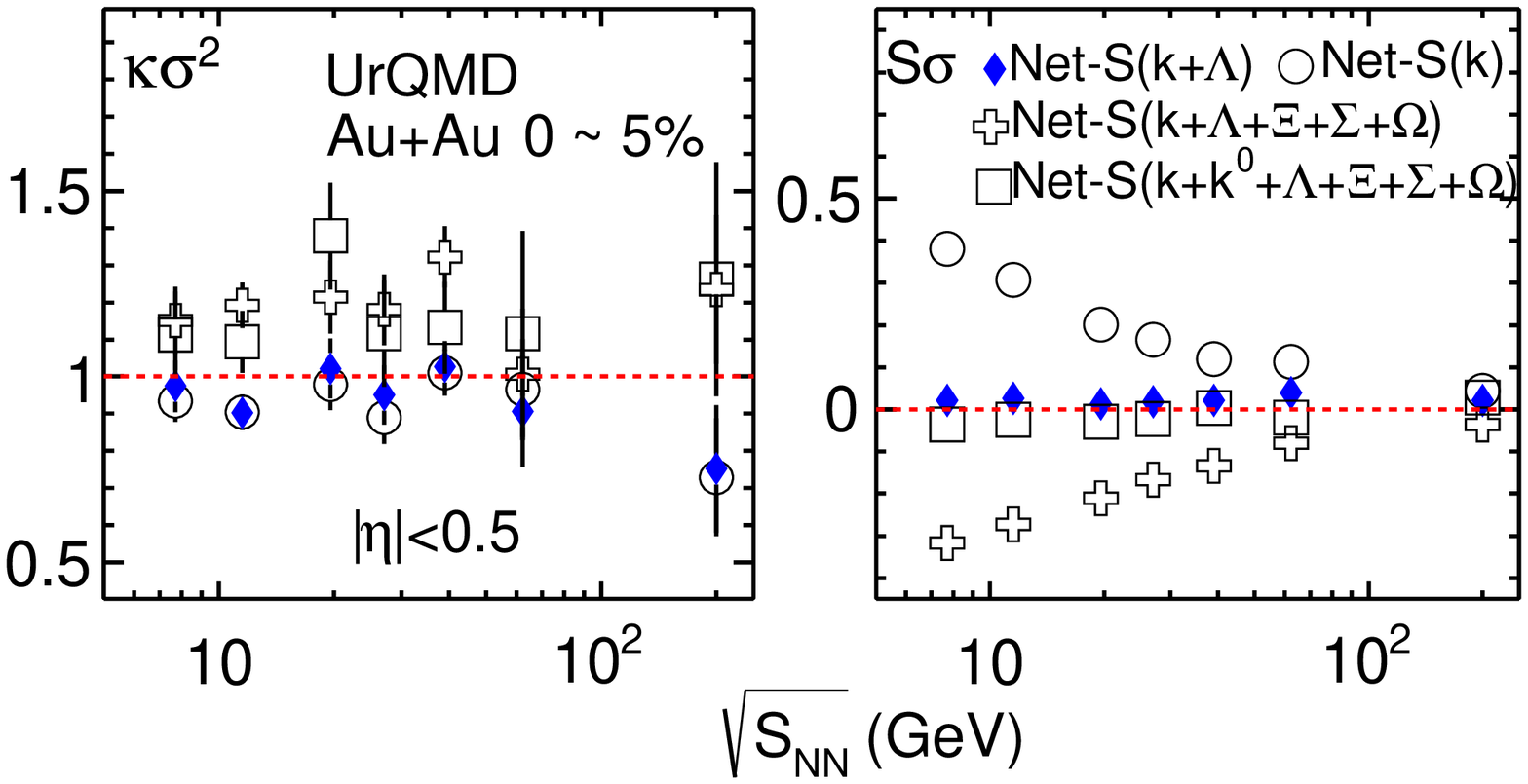}
    \caption{(color online) Energy dependence of cumulant ratios ($\kappa\sigma^{2},S\sigma$) of net-strangeness multiplicity distribution in the most central (0-5\%) Au+Au collisions at mid-rapidity ($|\eta|<0.5$) with UrQMD model.} 
   \label{fig:10}
  \end{center}
\end{figure*}

Figure~\ref{fig:5} displays various cumulants as a function of collision energy at mid-rapidity for the most central (0-5\%) Au+Au collisions from UrQMD model. 
We can observe that the even order cumulants ($C_{2},C_{4}$) increase with increasing the collision energy. However, the odd order cumulants ($C_{1},C_{3}$) of net-kaon decrease with increasing the collision energy. Additionally, the mean value of net-strangeness from $K+\Lambda$ and $K+K^{0}+\Lambda+\Xi+\Sigma+\Omega$ are close to zero. To understand those energy dependence trends, let's introduce the important properties of cumulants and moments~\cite{Stephanov:2008qz}. 
We use $C_{n}$ to denote the $n^{th}$ order cumulant of the probability distribution of the random variable $X$. 
According to the additivity of cumulants for independent variables, the additivity of cumulants can be written as:
\begin{eqnarray}
C_{n}(X+Y)=C_{n}(X)+C_{n}(Y)\text{,}
\end{eqnarray}
, where $X, Y$ are independent random variables, respectively. With the homogeneity properties of cumulants, we have
\begin{eqnarray}
C_{n}(X-Y)&=&C_{n}(X)+C_{n}(-Y) \\
&=&C_{n}(X)+(-1)^{n}C_{n}(Y)\text{.}
\end{eqnarray}

If the random variables $X$ and $Y$ are independent distributed as Poisson distributions, then the $X-Y$ will distributed as Skellam distribution, the cumulants of net-strangeness multiplicity distributions can be denoted by:
\begin{eqnarray}
C_{n}(X-Y)&=&C_{n}(X)+(-1)^{n}C_{n}(Y) \\
&=&<X>+(-1)^{n}<Y>
\end{eqnarray}
For odd order cumulants:
\begin{eqnarray}\label{C}
C_{1}(X-Y)=C_{3}(X-Y)=<X>-<Y>
\end{eqnarray}
For even order cumulants,we have:
\begin{eqnarray}\label{D}
C_{2}(X-Y)=C_{4}(X-Y)=<X>+<Y>
\end{eqnarray}
where, the $X$ denotes the number of anti-strange quark ($X=N_{\bar s}$), $Y$ is the number of strange quark($Y=N_{s}$) and $X-Y$ represents the net-strangeness number ($X-Y=N_{\bar s}-N_{s}$).  The energy dependence shown in Fig.~\ref{fig:5} can be attributed to the interplay between production mechanism of strange and anti-strange particles as a function of energy. At low energies, the associate production channel $NN \rightarrow N\Lambda K^{+}$ dominate the production of $K^{+}$, which makes the yield of $K^{+}$ larger than the yield of $K^{-}$. At high energies, due to pair production, the yields of the strange and anti-strange particles are very close to each other. For the case of net-kaon cumulants,  from Eq.~(\ref{C}), one can infer that the difference between odd order cumulants of $K^{+}$ and $K^{-}$ will become small as increasing the collision energies. Because of the additivity of the even order cumulants from $\bar{s}$ quark and $s$ quark as displayed by the Eq.~(\ref{D}), the even cumulants of net-strangeness show an increasing trend with increasing of the collision energy for different cases. Since more strange particles are included, we observe larger values of the even order cumulants. Meanwhile, the net-strangeness obtained from $K+K^{0}+\Lambda+\Xi+\Sigma+\Omega$ is a good approximation of the real net-strangeness and the values of odd order cumulants are close to zero. 

Figure~\ref{fig:6} and~\ref{fig:7} show $\kappa\sigma^{2}$ and $S\sigma$ of the net-strangeness distributions as a function of the average number of participant nucleons ($N_{part}$) in Au+Au collisions at $\sqrt{s_{NN}}$=7.7 to 200 GeV from UrQMD model. 
The $\kappa\sigma^{2}$ from different cases show weak centrality dependence. For the case of $K$ and $K+\Lambda$, the values of $\kappa\sigma^{2}$ are consistent with unity within errors. By including more multi-strange baryons, such as the case of $K+\Lambda+\Xi+\Sigma+\Omega$, the values of $\kappa\sigma^{2}$ are above unity. It indicates that the multi-strange baryons play an important role in the high order fluctuations of net-strangeness. It is similar as the two charged particles in net-charge fluctuations. The $S \sigma$ of net-kaon increase with increasing the number of participants and the values from $K+\Lambda+\Sigma+\Xi+\Omega$ and $K+K^{0}+\Lambda+\Sigma+\Xi+\Omega$ are negative. This can be explained by the Eq.~\ref{ks}. Due to the $C_{3}$ of net-strangeness are negative, the $S\sigma$ are negative.

Figure~\ref{fig:10} shows $\kappa\sigma^{2}$ and $S\sigma$ of net-strangeness distributions as a function of colliding energy for the most central (0-5\%) Au+Au collisions at mid-rapidity. The $\kappa\sigma^{2}$ of net-strangeness especially from $K$ and $K+\Lambda$ are closed to unity and show weak dependence on collision energy  If the multi-strange baryons are included in the calculations,  the values of $\kappa\sigma^{2}$ are above unity. We also observed that the $S\sigma$ of net-kaon distributions decrease with increasing collision energy and the values calculated from $K+\Lambda$ and $K+K^{0}+\Lambda+\Sigma+\Xi+\Omega$ are close to zero. One can observe different energy dependence behavior between $\kappa\sigma^{2}$ and $S\sigma$. It is because the skewness is sensitive to the asymmetry between strangeness and anti-strangeness while the kurtosis is sensitive to the multi-strange baryon with strangeness number $|s|>=2$. This is similar with the net-charge case that the charged two particles have strong effects on net-charge fluctuations~\cite{Karsch:2010ck}. If we take out the $K^{0}$, the values of the $S\sigma$ become negative and monotonically decrease with decreasing energy. It indicates that the neutral kaons carry a similar amount of strangeness than the charged kaons and show similar trends in the $S\sigma$ as a function of energy. On the other hand, based a hadronic transport model (JAM) model study in Au+Au collisions at 5 GeV, we find that the effects of hadronic scattering on proton fluctuations is negligible. One could also expect that the hadronic re-scattering effects are also small in net-kaon fluctuations, however, detail model studies are needed to carry out in the future.

\section{Summary}
We have performed systematical studies on the centrality, rapidity and energy dependence of the cumulants($C_{1}$ - $C_{4}$) and cumulant ratios ($\kappa\sigma^{2}$ and $S\sigma$) of net-strangeness distributions in Au+Au collisions at $\sqrt{s_{NN}}$=7.7, 11.5, 19.6, 27, 39, 62.4 and 200 GeV from UrQMD model. It is found that fluctuations of net-strangeness can be influenced by the production mechanism of strangeness as a function of collision energy which cause different results between lower energies and higher energies. Those difference can be understood as the associate production of $K^{+}$ play an important role at lower energies whereas pair production of strangeness and anti-strangeness dominates at higher energies. On the other hand,  our results show that $\kappa\sigma^{2}$ of net-strangeness have weak centrality and energy dependence.  In the current model study, we showed that the fraction of total strangeness carried in kaons are smaller than 45\% and monotonically decrease with decreasing energy. By comparing with the net-kaon fluctuations, we found that the multi-strange baryons play an important role in the fluctuations of net-strangeness. Those multi-strange baryons lead to the values of $\kappa\sigma^{2}$ become above unity. However, in terms of searching for non-monotonic energy dependence of the fluctuation observable near QCD critical point, the net-kaon fluctuations should still have sensitivity.  Since there has no QCD critical point and phase transition physics implemented in the UrQMD model, our model calculations can provide a baseline to search for the QCD critical point in heavy ion collisions.

\section*{Acknowledgement}
The work was supported in part by the MoST of China 973-Project No.2015CB856901, NSFC under grant No. 11575069 and 11221504.  

\bibliography{draft_ini}

\begin{thebibliography}{48}%
\makeatletter
\providecommand \@ifxundefined [1]{%
 \@ifx{#1\undefined}
}%
\providecommand \@ifnum [1]{%
 \ifnum #1\expandafter \@firstoftwo
 \else \expandafter \@secondoftwo
 \fi
}%
\providecommand \@ifx [1]{%
 \ifx #1\expandafter \@firstoftwo
 \else \expandafter \@secondoftwo
 \fi
}%
\providecommand \natexlab [1]{#1}%
\providecommand \enquote  [1]{``#1''}%
\providecommand \bibnamefont  [1]{#1}%
\providecommand \bibfnamefont [1]{#1}%
\providecommand \citenamefont [1]{#1}%
\providecommand \href@noop [0]{\@secondoftwo}%
\providecommand \href [0]{\begingroup \@sanitize@url \@href}%
\providecommand \@href[1]{\@@startlink{#1}\@@href}%
\providecommand \@@href[1]{\endgroup#1\@@endlink}%
\providecommand \@sanitize@url [0]{\catcode `\\12\catcode `\$12\catcode
  `\&12\catcode `\#12\catcode `\^12\catcode `\_12\catcode `\%12\relax}%
\providecommand \@@startlink[1]{}%
\providecommand \@@endlink[0]{}%
\providecommand \url  [0]{\begingroup\@sanitize@url \@url }%
\providecommand \@url [1]{\endgroup\@href {#1}{\urlprefix }}%
\providecommand \urlprefix  [0]{URL }%
\providecommand \Eprint [0]{\href }%
\providecommand \doibase [0]{http://dx.doi.org/}%
\providecommand \selectlanguage [0]{\@gobble}%
\providecommand \bibinfo  [0]{\@secondoftwo}%
\providecommand \bibfield  [0]{\@secondoftwo}%
\providecommand \translation [1]{[#1]}%
\providecommand \BibitemOpen [0]{}%
\providecommand \bibitemStop [0]{}%
\providecommand \bibitemNoStop [0]{.\EOS\space}%
\providecommand \EOS [0]{\spacefactor3000\relax}%
\providecommand \BibitemShut  [1]{\csname bibitem#1\endcsname}%
\let\auto@bib@innerbib\@empty
\bibitem [{\citenamefont {Aoki}\ \emph
  {et~al.}(2006{\natexlab{a}})\citenamefont {Aoki}, \citenamefont {Endrodi},
  \citenamefont {Fodor}, \citenamefont {Katz},\ and\ \citenamefont
  {Szabo}}]{Aoki:2006we}%
  \BibitemOpen
  \bibfield  {author} {\bibinfo {author} {\bibfnamefont {Y.}~\bibnamefont
  {Aoki}}, \bibinfo {author} {\bibfnamefont {G.}~\bibnamefont {Endrodi}},
  \bibinfo {author} {\bibfnamefont {Z.}~\bibnamefont {Fodor}}, \bibinfo
  {author} {\bibfnamefont {S.~D.}\ \bibnamefont {Katz}}, \ and\ \bibinfo
  {author} {\bibfnamefont {K.~K.}\ \bibnamefont {Szabo}},\ }\href {\doibase
  10.1038/nature05120} {\bibfield  {journal} {\bibinfo  {journal} {Nature}\
  }\textbf {\bibinfo {volume} {443}},\ \bibinfo {pages} {675} (\bibinfo {year}
  {2006}{\natexlab{a}})},\ \Eprint {http://arxiv.org/abs/hep-lat/0611014}
  {arXiv:hep-lat/0611014 [hep-lat]} \BibitemShut {NoStop}%
\bibitem [{\citenamefont {Aoki}\ \emph
  {et~al.}(2006{\natexlab{b}})\citenamefont {Aoki}, \citenamefont {Fodor},
  \citenamefont {Katz},\ and\ \citenamefont {Szabo}}]{Aoki:2006br}%
  \BibitemOpen
  \bibfield  {author} {\bibinfo {author} {\bibfnamefont {Y.}~\bibnamefont
  {Aoki}}, \bibinfo {author} {\bibfnamefont {Z.}~\bibnamefont {Fodor}},
  \bibinfo {author} {\bibfnamefont {S.~D.}\ \bibnamefont {Katz}}, \ and\
  \bibinfo {author} {\bibfnamefont {K.~K.}\ \bibnamefont {Szabo}},\ }\href
  {\doibase 10.1016/j.physletb.2006.10.021} {\bibfield  {journal} {\bibinfo
  {journal} {Phys. Lett.}\ }\textbf {\bibinfo {volume} {B643}},\ \bibinfo
  {pages} {46} (\bibinfo {year} {2006}{\natexlab{b}})},\ \Eprint
  {http://arxiv.org/abs/hep-lat/0609068} {arXiv:hep-lat/0609068 [hep-lat]}
  \BibitemShut {NoStop}%
\bibitem [{\citenamefont {Ejiri}(2008)}]{Ejiri:2008xt}%
  \BibitemOpen
  \bibfield  {author} {\bibinfo {author} {\bibfnamefont {S.}~\bibnamefont
  {Ejiri}},\ }\href {\doibase 10.1103/PhysRevD.78.074507} {\bibfield  {journal}
  {\bibinfo  {journal} {Phys. Rev.}\ }\textbf {\bibinfo {volume} {D78}},\
  \bibinfo {pages} {074507} (\bibinfo {year} {2008})},\ \Eprint
  {http://arxiv.org/abs/0804.3227} {arXiv:0804.3227 [hep-lat]} \BibitemShut
  {NoStop}%
\bibitem [{\citenamefont {Endrodi}\ \emph {et~al.}(2011)\citenamefont
  {Endrodi}, \citenamefont {Fodor}, \citenamefont {Katz},\ and\ \citenamefont
  {Szabo}}]{Endrodi:2011gv}%
  \BibitemOpen
  \bibfield  {author} {\bibinfo {author} {\bibfnamefont {G.}~\bibnamefont
  {Endrodi}}, \bibinfo {author} {\bibfnamefont {Z.}~\bibnamefont {Fodor}},
  \bibinfo {author} {\bibfnamefont {S.~D.}\ \bibnamefont {Katz}}, \ and\
  \bibinfo {author} {\bibfnamefont {K.~K.}\ \bibnamefont {Szabo}},\ }\href
  {\doibase 10.1007/JHEP04(2011)001} {\bibfield  {journal} {\bibinfo  {journal}
  {JHEP}\ }\textbf {\bibinfo {volume} {04}},\ \bibinfo {pages} {001} (\bibinfo
  {year} {2011})},\ \Eprint {http://arxiv.org/abs/1102.1356} {arXiv:1102.1356
  [hep-lat]} \BibitemShut {NoStop}%
\bibitem [{\citenamefont {de~Forcrand}\ and\ \citenamefont
  {Philipsen}(2002)}]{deForcrand:2002hgr}%
  \BibitemOpen
  \bibfield  {author} {\bibinfo {author} {\bibfnamefont {P.}~\bibnamefont
  {de~Forcrand}}\ and\ \bibinfo {author} {\bibfnamefont {O.}~\bibnamefont
  {Philipsen}},\ }\href {\doibase 10.1016/S0550-3213(02)00626-0} {\bibfield
  {journal} {\bibinfo  {journal} {Nucl. Phys.}\ }\textbf {\bibinfo {volume}
  {B642}},\ \bibinfo {pages} {290} (\bibinfo {year} {2002})},\ \Eprint
  {http://arxiv.org/abs/hep-lat/0205016} {arXiv:hep-lat/0205016 [hep-lat]}
  \BibitemShut {NoStop}%
\bibitem [{\citenamefont {Stephanov}(2004)}]{Stephanov:2004wx}%
  \BibitemOpen
  \bibfield  {author} {\bibinfo {author} {\bibfnamefont {M.~A.}\ \bibnamefont
  {Stephanov}},\ }\href {\doibase 10.1142/S0217751X05027965} {\bibfield
  {journal} {\bibinfo  {journal} {Prog. Theor. Phys. Suppl.}\ }\textbf
  {\bibinfo {volume} {153}},\ \bibinfo {pages} {139} (\bibinfo {year}
  {2004})},\ \bibinfo {note} {[Int. J. Mod. Phys.A20,4387(2005)]},\ \Eprint
  {http://arxiv.org/abs/hep-ph/0402115} {arXiv:hep-ph/0402115 [hep-ph]}
  \BibitemShut {NoStop}%
\bibitem [{\citenamefont {Fodor}\ and\ \citenamefont
  {Katz}(2004)}]{Fodor:2004nz}%
  \BibitemOpen
  \bibfield  {author} {\bibinfo {author} {\bibfnamefont {Z.}~\bibnamefont
  {Fodor}}\ and\ \bibinfo {author} {\bibfnamefont {S.~D.}\ \bibnamefont
  {Katz}},\ }\href {\doibase 10.1088/1126-6708/2004/04/050} {\bibfield
  {journal} {\bibinfo  {journal} {JHEP}\ }\textbf {\bibinfo {volume} {04}},\
  \bibinfo {pages} {050} (\bibinfo {year} {2004})},\ \Eprint
  {http://arxiv.org/abs/hep-lat/0402006} {arXiv:hep-lat/0402006 [hep-lat]}
  \BibitemShut {NoStop}%
\bibitem [{\citenamefont {Stephanov}(2009)}]{Stephanov:2008qz}%
  \BibitemOpen
  \bibfield  {author} {\bibinfo {author} {\bibfnamefont {M.~A.}\ \bibnamefont
  {Stephanov}},\ }\href {\doibase 10.1103/PhysRevLett.102.032301} {\bibfield
  {journal} {\bibinfo  {journal} {Phys. Rev. Lett.}\ }\textbf {\bibinfo
  {volume} {102}},\ \bibinfo {pages} {032301} (\bibinfo {year} {2009})},\
  \Eprint {http://arxiv.org/abs/0809.3450} {arXiv:0809.3450 [hep-ph]}
  \BibitemShut {NoStop}%
\bibitem [{\citenamefont {Athanasiou}\ \emph {et~al.}(2010)\citenamefont
  {Athanasiou}, \citenamefont {Rajagopal},\ and\ \citenamefont
  {Stephanov}}]{Athanasiou:2010kw}%
  \BibitemOpen
  \bibfield  {author} {\bibinfo {author} {\bibfnamefont {C.}~\bibnamefont
  {Athanasiou}}, \bibinfo {author} {\bibfnamefont {K.}~\bibnamefont
  {Rajagopal}}, \ and\ \bibinfo {author} {\bibfnamefont {M.}~\bibnamefont
  {Stephanov}},\ }\href {\doibase 10.1103/PhysRevD.82.074008} {\bibfield
  {journal} {\bibinfo  {journal} {Phys. Rev.}\ }\textbf {\bibinfo {volume}
  {D82}},\ \bibinfo {pages} {074008} (\bibinfo {year} {2010})},\ \Eprint
  {http://arxiv.org/abs/1006.4636} {arXiv:1006.4636 [hep-ph]} \BibitemShut
  {NoStop}%
\bibitem [{\citenamefont {Hatta}\ and\ \citenamefont
  {Stephanov}(2003)}]{Hatta:2003wn}%
  \BibitemOpen
  \bibfield  {author} {\bibinfo {author} {\bibfnamefont {Y.}~\bibnamefont
  {Hatta}}\ and\ \bibinfo {author} {\bibfnamefont {M.~A.}\ \bibnamefont
  {Stephanov}},\ }\href {\doibase 10.1103/PhysRevLett.91.102003,
  10.1103/PhysRevLett.91.129901} {\bibfield  {journal} {\bibinfo  {journal}
  {Phys. Rev. Lett.}\ }\textbf {\bibinfo {volume} {91}},\ \bibinfo {pages}
  {102003} (\bibinfo {year} {2003})},\ \bibinfo {note} {[Erratum: Phys. Rev.
  Lett.91,129901(2003)]},\ \Eprint {http://arxiv.org/abs/hep-ph/0302002}
  {arXiv:hep-ph/0302002 [hep-ph]} \BibitemShut {NoStop}%
\bibitem [{\citenamefont {Gavai}\ and\ \citenamefont
  {Gupta}(2011)}]{Gavai:2010zn}%
  \BibitemOpen
  \bibfield  {author} {\bibinfo {author} {\bibfnamefont {R.~V.}\ \bibnamefont
  {Gavai}}\ and\ \bibinfo {author} {\bibfnamefont {S.}~\bibnamefont {Gupta}},\
  }\href {\doibase 10.1016/j.physletb.2011.01.006} {\bibfield  {journal}
  {\bibinfo  {journal} {Phys. Lett.}\ }\textbf {\bibinfo {volume} {B696}},\
  \bibinfo {pages} {459} (\bibinfo {year} {2011})},\ \Eprint
  {http://arxiv.org/abs/1001.3796} {arXiv:1001.3796 [hep-lat]} \BibitemShut
  {NoStop}%
\bibitem [{\citenamefont {Gavai}\ and\ \citenamefont
  {Gupta}(2008)}]{Gavai:2008zr}%
  \BibitemOpen
  \bibfield  {author} {\bibinfo {author} {\bibfnamefont {R.~V.}\ \bibnamefont
  {Gavai}}\ and\ \bibinfo {author} {\bibfnamefont {S.}~\bibnamefont {Gupta}},\
  }\href {\doibase 10.1103/PhysRevD.78.114503} {\bibfield  {journal} {\bibinfo
  {journal} {Phys. Rev.}\ }\textbf {\bibinfo {volume} {D78}},\ \bibinfo {pages}
  {114503} (\bibinfo {year} {2008})},\ \Eprint {http://arxiv.org/abs/0806.2233}
  {arXiv:0806.2233 [hep-lat]} \BibitemShut {NoStop}%
\bibitem [{\citenamefont {Cheng}\ \emph {et~al.}(2009)\citenamefont {Cheng}
  \emph {et~al.}}]{Cheng:2008zh}%
  \BibitemOpen
  \bibfield  {author} {\bibinfo {author} {\bibfnamefont {M.}~\bibnamefont
  {Cheng}} \emph {et~al.},\ }\href {\doibase 10.1103/PhysRevD.79.074505}
  {\bibfield  {journal} {\bibinfo  {journal} {Phys. Rev.}\ }\textbf {\bibinfo
  {volume} {D79}},\ \bibinfo {pages} {074505} (\bibinfo {year} {2009})},\
  \Eprint {http://arxiv.org/abs/0811.1006} {arXiv:0811.1006 [hep-lat]}
  \BibitemShut {NoStop}%
\bibitem [{\citenamefont {Bazavov}\ \emph
  {et~al.}(2012{\natexlab{a}})\citenamefont {Bazavov} \emph
  {et~al.}}]{Bazavov:2012vg}%
  \BibitemOpen
  \bibfield  {author} {\bibinfo {author} {\bibfnamefont {A.}~\bibnamefont
  {Bazavov}} \emph {et~al.},\ }\href {\doibase 10.1103/PhysRevLett.109.192302}
  {\bibfield  {journal} {\bibinfo  {journal} {Phys. Rev. Lett.}\ }\textbf
  {\bibinfo {volume} {109}},\ \bibinfo {pages} {192302} (\bibinfo {year}
  {2012}{\natexlab{a}})},\ \Eprint {http://arxiv.org/abs/1208.1220}
  {arXiv:1208.1220 [hep-lat]} \BibitemShut {NoStop}%
\bibitem [{\citenamefont {Ding}\ \emph {et~al.}(2015)\citenamefont {Ding},
  \citenamefont {Karsch},\ and\ \citenamefont {Mukherjee}}]{Ding:2015ona}%
  \BibitemOpen
  \bibfield  {author} {\bibinfo {author} {\bibfnamefont {H.-T.}\ \bibnamefont
  {Ding}}, \bibinfo {author} {\bibfnamefont {F.}~\bibnamefont {Karsch}}, \ and\
  \bibinfo {author} {\bibfnamefont {S.}~\bibnamefont {Mukherjee}},\ }\href
  {\doibase 10.1142/S0218301315300076} {\bibfield  {journal} {\bibinfo
  {journal} {Int. J. Mod. Phys.}\ }\textbf {\bibinfo {volume} {E24}},\ \bibinfo
  {pages} {1530007} (\bibinfo {year} {2015})},\ \Eprint
  {http://arxiv.org/abs/1504.05274} {arXiv:1504.05274 [hep-lat]} \BibitemShut
  {NoStop}%
\bibitem [{\citenamefont {Bazavov}\ \emph
  {et~al.}(2012{\natexlab{b}})\citenamefont {Bazavov} \emph
  {et~al.}}]{Bazavov:2012jq}%
  \BibitemOpen
  \bibfield  {author} {\bibinfo {author} {\bibfnamefont {A.}~\bibnamefont
  {Bazavov}} \emph {et~al.} (\bibinfo {collaboration} {HotQCD}),\ }\href
  {\doibase 10.1103/PhysRevD.86.034509} {\bibfield  {journal} {\bibinfo
  {journal} {Phys. Rev.}\ }\textbf {\bibinfo {volume} {D86}},\ \bibinfo {pages}
  {034509} (\bibinfo {year} {2012}{\natexlab{b}})},\ \Eprint
  {http://arxiv.org/abs/1203.0784} {arXiv:1203.0784 [hep-lat]} \BibitemShut
  {NoStop}%
\bibitem [{\citenamefont {Friman}\ \emph {et~al.}(2011)\citenamefont {Friman},
  \citenamefont {Karsch}, \citenamefont {Redlich},\ and\ \citenamefont
  {Skokov}}]{Friman:2011pf}%
  \BibitemOpen
  \bibfield  {author} {\bibinfo {author} {\bibfnamefont {B.}~\bibnamefont
  {Friman}}, \bibinfo {author} {\bibfnamefont {F.}~\bibnamefont {Karsch}},
  \bibinfo {author} {\bibfnamefont {K.}~\bibnamefont {Redlich}}, \ and\
  \bibinfo {author} {\bibfnamefont {V.}~\bibnamefont {Skokov}},\ }\href
  {\doibase 10.1140/epjc/s10052-011-1694-2} {\bibfield  {journal} {\bibinfo
  {journal} {Eur. Phys. J.}\ }\textbf {\bibinfo {volume} {C71}},\ \bibinfo
  {pages} {1694} (\bibinfo {year} {2011})},\ \Eprint
  {http://arxiv.org/abs/1103.3511} {arXiv:1103.3511 [hep-ph]} \BibitemShut
  {NoStop}%
\bibitem [{\citenamefont {Mukherjee}\ \emph {et~al.}(2015)\citenamefont
  {Mukherjee}, \citenamefont {Venugopalan},\ and\ \citenamefont
  {Yin}}]{Mukherjee:2015swa}%
  \BibitemOpen
  \bibfield  {author} {\bibinfo {author} {\bibfnamefont {S.}~\bibnamefont
  {Mukherjee}}, \bibinfo {author} {\bibfnamefont {R.}~\bibnamefont
  {Venugopalan}}, \ and\ \bibinfo {author} {\bibfnamefont {Y.}~\bibnamefont
  {Yin}},\ }\href {\doibase 10.1103/PhysRevC.92.034912} {\bibfield  {journal}
  {\bibinfo  {journal} {Phys. Rev.}\ }\textbf {\bibinfo {volume} {C92}},\
  \bibinfo {pages} {034912} (\bibinfo {year} {2015})},\ \Eprint
  {http://arxiv.org/abs/1506.00645} {arXiv:1506.00645 [hep-ph]} \BibitemShut
  {NoStop}%
\bibitem [{\citenamefont {Morita}\ \emph {et~al.}(2015)\citenamefont {Morita},
  \citenamefont {Friman},\ and\ \citenamefont {Redlich}}]{Morita:2014fda}%
  \BibitemOpen
  \bibfield  {author} {\bibinfo {author} {\bibfnamefont {K.}~\bibnamefont
  {Morita}}, \bibinfo {author} {\bibfnamefont {B.}~\bibnamefont {Friman}}, \
  and\ \bibinfo {author} {\bibfnamefont {K.}~\bibnamefont {Redlich}},\ }\href
  {\doibase 10.1016/j.physletb.2014.12.037} {\bibfield  {journal} {\bibinfo
  {journal} {Phys. Lett.}\ }\textbf {\bibinfo {volume} {B741}},\ \bibinfo
  {pages} {178} (\bibinfo {year} {2015})},\ \Eprint
  {http://arxiv.org/abs/1402.5982} {arXiv:1402.5982 [hep-ph]} \BibitemShut
  {NoStop}%
\bibitem [{\citenamefont {Karsch}\ and\ \citenamefont
  {Redlich}(2011)}]{Karsch:2010ck}%
  \BibitemOpen
  \bibfield  {author} {\bibinfo {author} {\bibfnamefont {F.}~\bibnamefont
  {Karsch}}\ and\ \bibinfo {author} {\bibfnamefont {K.}~\bibnamefont
  {Redlich}},\ }\href {\doibase 10.1016/j.physletb.2010.10.046} {\bibfield
  {journal} {\bibinfo  {journal} {Phys. Lett.}\ }\textbf {\bibinfo {volume}
  {B695}},\ \bibinfo {pages} {136} (\bibinfo {year} {2011})},\ \Eprint
  {http://arxiv.org/abs/1007.2581} {arXiv:1007.2581 [hep-ph]} \BibitemShut
  {NoStop}%
\bibitem [{\citenamefont {Garg}\ \emph {et~al.}(2013)\citenamefont {Garg},
  \citenamefont {Mishra}, \citenamefont {Netrakanti}, \citenamefont {Mohanty},
  \citenamefont {Mohanty}, \citenamefont {Singh},\ and\ \citenamefont
  {Xu}}]{Garg:2013ata}%
  \BibitemOpen
  \bibfield  {author} {\bibinfo {author} {\bibfnamefont {P.}~\bibnamefont
  {Garg}}, \bibinfo {author} {\bibfnamefont {D.~K.}\ \bibnamefont {Mishra}},
  \bibinfo {author} {\bibfnamefont {P.~K.}\ \bibnamefont {Netrakanti}},
  \bibinfo {author} {\bibfnamefont {B.}~\bibnamefont {Mohanty}}, \bibinfo
  {author} {\bibfnamefont {A.~K.}\ \bibnamefont {Mohanty}}, \bibinfo {author}
  {\bibfnamefont {B.~K.}\ \bibnamefont {Singh}}, \ and\ \bibinfo {author}
  {\bibfnamefont {N.}~\bibnamefont {Xu}},\ }\href {\doibase
  10.1016/j.physletb.2013.09.019} {\bibfield  {journal} {\bibinfo  {journal}
  {Phys. Lett.}\ }\textbf {\bibinfo {volume} {B726}},\ \bibinfo {pages} {691}
  (\bibinfo {year} {2013})},\ \Eprint {http://arxiv.org/abs/1304.7133}
  {arXiv:1304.7133 [nucl-ex]} \BibitemShut {NoStop}%
\bibitem [{\citenamefont {Fu}(2013)}]{Fu:2013gga}%
  \BibitemOpen
  \bibfield  {author} {\bibinfo {author} {\bibfnamefont {J.}~\bibnamefont
  {Fu}},\ }\href {\doibase 10.1016/j.physletb.2013.04.018} {\bibfield
  {journal} {\bibinfo  {journal} {Phys. Lett.}\ }\textbf {\bibinfo {volume}
  {B722}},\ \bibinfo {pages} {144} (\bibinfo {year} {2013})}\BibitemShut
  {NoStop}%
\bibitem [{\citenamefont {Nahrgang}\ \emph {et~al.}(2015)\citenamefont
  {Nahrgang}, \citenamefont {Bluhm}, \citenamefont {Alba}, \citenamefont
  {Bellwied},\ and\ \citenamefont {Ratti}}]{Nahrgang:2014fza}%
  \BibitemOpen
  \bibfield  {author} {\bibinfo {author} {\bibfnamefont {M.}~\bibnamefont
  {Nahrgang}}, \bibinfo {author} {\bibfnamefont {M.}~\bibnamefont {Bluhm}},
  \bibinfo {author} {\bibfnamefont {P.}~\bibnamefont {Alba}}, \bibinfo {author}
  {\bibfnamefont {R.}~\bibnamefont {Bellwied}}, \ and\ \bibinfo {author}
  {\bibfnamefont {C.}~\bibnamefont {Ratti}},\ }\href {\doibase
  10.1140/epjc/s10052-015-3775-0} {\bibfield  {journal} {\bibinfo  {journal}
  {Eur. Phys. J.}\ }\textbf {\bibinfo {volume} {C75}},\ \bibinfo {pages} {573}
  (\bibinfo {year} {2015})},\ \Eprint {http://arxiv.org/abs/1402.1238}
  {arXiv:1402.1238 [hep-ph]} \BibitemShut {NoStop}%
\bibitem [{\citenamefont {Luo}\ \emph {et~al.}(2009)\citenamefont {Luo},
  \citenamefont {Shao}, \citenamefont {Li},\ and\ \citenamefont
  {Chen}}]{Luo:2009sx}%
  \BibitemOpen
  \bibfield  {author} {\bibinfo {author} {\bibfnamefont {X.}~\bibnamefont
  {Luo}}, \bibinfo {author} {\bibfnamefont {M.}~\bibnamefont {Shao}}, \bibinfo
  {author} {\bibfnamefont {C.}~\bibnamefont {Li}}, \ and\ \bibinfo {author}
  {\bibfnamefont {H.}~\bibnamefont {Chen}},\ }\href {\doibase
  10.1016/j.physletb.2009.02.044} {\bibfield  {journal} {\bibinfo  {journal}
  {Phys. Lett.}\ }\textbf {\bibinfo {volume} {B673}},\ \bibinfo {pages} {268}
  (\bibinfo {year} {2009})},\ \Eprint {http://arxiv.org/abs/0903.0024}
  {arXiv:0903.0024 [nucl-th]} \BibitemShut {NoStop}%
\bibitem [{\citenamefont {Nahrgang}(2015)}]{Nahrgang:2015tva}%
  \BibitemOpen
  \bibfield  {author} {\bibinfo {author} {\bibfnamefont {M.}~\bibnamefont
  {Nahrgang}},\ }\href@noop {} {\bibfield  {journal} {\bibinfo  {journal}
  {PoS}\ }\textbf {\bibinfo {volume} {CPOD2014}},\ \bibinfo {pages} {032}
  (\bibinfo {year} {2015})},\ \Eprint {http://arxiv.org/abs/1510.08146}
  {arXiv:1510.08146 [nucl-th]} \BibitemShut {NoStop}%
\bibitem [{\citenamefont {Chen}\ \emph {et~al.}(2015)\citenamefont {Chen},
  \citenamefont {Deng},\ and\ \citenamefont {Labun}}]{Chen:2014ufa}%
  \BibitemOpen
  \bibfield  {author} {\bibinfo {author} {\bibfnamefont {J.-W.}\ \bibnamefont
  {Chen}}, \bibinfo {author} {\bibfnamefont {J.}~\bibnamefont {Deng}}, \ and\
  \bibinfo {author} {\bibfnamefont {L.}~\bibnamefont {Labun}},\ }\href
  {\doibase 10.1103/PhysRevD.92.054019} {\bibfield  {journal} {\bibinfo
  {journal} {Phys. Rev.}\ }\textbf {\bibinfo {volume} {D92}},\ \bibinfo {pages}
  {054019} (\bibinfo {year} {2015})},\ \Eprint {http://arxiv.org/abs/1410.5454}
  {arXiv:1410.5454 [hep-ph]} \BibitemShut {NoStop}%
\bibitem [{\citenamefont {Jiang}\ \emph {et~al.}(2016)\citenamefont {Jiang},
  \citenamefont {Li},\ and\ \citenamefont {Song}}]{Jiang:2015cnt}%
  \BibitemOpen
  \bibfield  {author} {\bibinfo {author} {\bibfnamefont {L.}~\bibnamefont
  {Jiang}}, \bibinfo {author} {\bibfnamefont {P.}~\bibnamefont {Li}}, \ and\
  \bibinfo {author} {\bibfnamefont {H.}~\bibnamefont {Song}},\ }\href {\doibase
  10.1016/j.nuclphysa.2016.01.034} {\bibfield  {journal} {\bibinfo  {journal}
  {Nucl. Phys.}\ }\textbf {\bibinfo {volume} {A956}},\ \bibinfo {pages} {360}
  (\bibinfo {year} {2016})},\ \Eprint {http://arxiv.org/abs/1512.07373}
  {arXiv:1512.07373 [nucl-th]} \BibitemShut {NoStop}%
\bibitem [{\citenamefont {Gupta}\ \emph {et~al.}(2011)\citenamefont {Gupta},
  \citenamefont {Luo}, \citenamefont {Mohanty}, \citenamefont {Ritter},\ and\
  \citenamefont {Xu}}]{Gupta:2011wh}%
  \BibitemOpen
  \bibfield  {author} {\bibinfo {author} {\bibfnamefont {S.}~\bibnamefont
  {Gupta}}, \bibinfo {author} {\bibfnamefont {X.}~\bibnamefont {Luo}}, \bibinfo
  {author} {\bibfnamefont {B.}~\bibnamefont {Mohanty}}, \bibinfo {author}
  {\bibfnamefont {H.~G.}\ \bibnamefont {Ritter}}, \ and\ \bibinfo {author}
  {\bibfnamefont {N.}~\bibnamefont {Xu}},\ }\href {\doibase
  10.1126/science.1204621} {\bibfield  {journal} {\bibinfo  {journal}
  {Science}\ }\textbf {\bibinfo {volume} {332}},\ \bibinfo {pages} {1525}
  (\bibinfo {year} {2011})},\ \Eprint {http://arxiv.org/abs/1105.3934}
  {arXiv:1105.3934 [hep-ph]} \BibitemShut {NoStop}%
\bibitem [{\citenamefont {Asakawa}\ \emph {et~al.}(2009)\citenamefont
  {Asakawa}, \citenamefont {Ejiri},\ and\ \citenamefont
  {Kitazawa}}]{Asakawa:2009aj}%
  \BibitemOpen
  \bibfield  {author} {\bibinfo {author} {\bibfnamefont {M.}~\bibnamefont
  {Asakawa}}, \bibinfo {author} {\bibfnamefont {S.}~\bibnamefont {Ejiri}}, \
  and\ \bibinfo {author} {\bibfnamefont {M.}~\bibnamefont {Kitazawa}},\ }\href
  {\doibase 10.1103/PhysRevLett.103.262301} {\bibfield  {journal} {\bibinfo
  {journal} {Phys. Rev. Lett.}\ }\textbf {\bibinfo {volume} {103}},\ \bibinfo
  {pages} {262301} (\bibinfo {year} {2009})},\ \Eprint
  {http://arxiv.org/abs/0904.2089} {arXiv:0904.2089 [nucl-th]} \BibitemShut
  {NoStop}%
\bibitem [{\citenamefont {Stephanov}(2011)}]{Stephanov:2011pb}%
  \BibitemOpen
  \bibfield  {author} {\bibinfo {author} {\bibfnamefont {M.~A.}\ \bibnamefont
  {Stephanov}},\ }\href {\doibase 10.1103/PhysRevLett.107.052301} {\bibfield
  {journal} {\bibinfo  {journal} {Phys. Rev. Lett.}\ }\textbf {\bibinfo
  {volume} {107}},\ \bibinfo {pages} {052301} (\bibinfo {year} {2011})},\
  \Eprint {http://arxiv.org/abs/1104.1627} {arXiv:1104.1627 [hep-ph]}
  \BibitemShut {NoStop}%
\bibitem [{\citenamefont {Thäder}(2016)}]{Thader:2016gpa}%
  \BibitemOpen
  \bibfield  {author} {\bibinfo {author} {\bibfnamefont {J.}~\bibnamefont
  {Thaeder}} (\bibinfo {collaboration} {STAR}),\ }\href {\doibase
  10.1016/j.nuclphysa.2016.02.047} {\bibfield  {journal} {\bibinfo  {journal}
  {Nucl. Phys.}\ }\textbf {\bibinfo {volume} {A956}},\ \bibinfo {pages} {320}
  (\bibinfo {year} {2016})},\ \Eprint {http://arxiv.org/abs/1601.00951}
  {arXiv:1601.00951 [nucl-ex]} \BibitemShut {NoStop}%
\bibitem [{\citenamefont {Al.}(2001)}]{Al2001An}%
  \BibitemOpen
  \bibfield  {author} {\bibinfo {author} {\bibfnamefont {E.}~\bibnamefont
  {Al.}},\ }\href@noop {} {\bibfield  {journal} {\bibinfo  {journal} {Journal
  of Physics G Nuclear \& Particle Physics}\ }\textbf {\bibinfo {volume}
  {27}},\ \bibinfo {pages} {311} (\bibinfo {year} {2001})}\BibitemShut
  {NoStop}%
\bibitem [{\citenamefont {Antinori}\ \emph {et~al.}(2004)\citenamefont
  {Antinori} \emph {et~al.}}]{Antinori:2004ee}%
  \BibitemOpen
  \bibfield  {author} {\bibinfo {author} {\bibfnamefont {F.}~\bibnamefont
  {Antinori}} \emph {et~al.} (\bibinfo {collaboration} {NA57}),\ }\href
  {\doibase 10.1016/j.physletb.2004.05.025} {\bibfield  {journal} {\bibinfo
  {journal} {Phys. Lett.}\ }\textbf {\bibinfo {volume} {B595}},\ \bibinfo
  {pages} {68} (\bibinfo {year} {2004})},\ \Eprint
  {http://arxiv.org/abs/nucl-ex/0403022} {arXiv:nucl-ex/0403022 [nucl-ex]}
  \BibitemShut {NoStop}%
\bibitem [{\citenamefont {Alt}\ \emph {et~al.}(2005)\citenamefont {Alt} \emph
  {et~al.}}]{Alt:2004kq}%
  \BibitemOpen
  \bibfield  {author} {\bibinfo {author} {\bibfnamefont {C.}~\bibnamefont
  {Alt}} \emph {et~al.} (\bibinfo {collaboration} {NA49}),\ }\href {\doibase
  10.1103/PhysRevLett.94.192301} {\bibfield  {journal} {\bibinfo  {journal}
  {Phys. Rev. Lett.}\ }\textbf {\bibinfo {volume} {94}},\ \bibinfo {pages}
  {192301} (\bibinfo {year} {2005})},\ \Eprint
  {http://arxiv.org/abs/nucl-ex/0409004} {arXiv:nucl-ex/0409004 [nucl-ex]}
  \BibitemShut {NoStop}%
\bibitem [{\citenamefont {Adler}\ \emph {et~al.}(2002)\citenamefont {Adler}
  \emph {et~al.}}]{Adler:2002uv}%
  \BibitemOpen
  \bibfield  {author} {\bibinfo {author} {\bibfnamefont {C.}~\bibnamefont
  {Adler}} \emph {et~al.} (\bibinfo {collaboration} {STAR}),\ }\href {\doibase
  10.1103/PhysRevLett.89.092301} {\bibfield  {journal} {\bibinfo  {journal}
  {Phys. Rev. Lett.}\ }\textbf {\bibinfo {volume} {89}},\ \bibinfo {pages}
  {092301} (\bibinfo {year} {2002})},\ \Eprint
  {http://arxiv.org/abs/nucl-ex/0203016} {arXiv:nucl-ex/0203016 [nucl-ex]}
  \BibitemShut {NoStop}%
\bibitem [{\citenamefont {Bleicher}\ \emph {et~al.}(1999)\citenamefont
  {Bleicher} \emph {et~al.}}]{Bleicher:1999xi}%
  \BibitemOpen
  \bibfield  {author} {\bibinfo {author} {\bibfnamefont {M.}~\bibnamefont
  {Bleicher}} \emph {et~al.},\ }\href {\doibase 10.1088/0954-3899/25/9/308}
  {\bibfield  {journal} {\bibinfo  {journal} {J. Phys.}\ }\textbf {\bibinfo
  {volume} {G25}},\ \bibinfo {pages} {1859} (\bibinfo {year} {1999})},\ \Eprint
  {http://arxiv.org/abs/hep-ph/9909407} {arXiv:hep-ph/9909407 [hep-ph]}
  \BibitemShut {NoStop}%
\bibitem [{\citenamefont {Bass}\ \emph {et~al.}(1998)\citenamefont {Bass} \emph
  {et~al.}}]{Bass:1998ca}%
  \BibitemOpen
  \bibfield  {author} {\bibinfo {author} {\bibfnamefont {S.~A.}\ \bibnamefont
  {Bass}} \emph {et~al.},\ }\href {\doibase 10.1016/S0146-6410(98)00058-1}
  {\bibfield  {journal} {\bibinfo  {journal} {Prog. Part. Nucl. Phys.}\
  }\textbf {\bibinfo {volume} {41}},\ \bibinfo {pages} {255} (\bibinfo {year}
  {1998})},\ \bibinfo {note} {[Prog. Part. Nucl. Phys.41,225(1998)]},\ \Eprint
  {http://arxiv.org/abs/nucl-th/9803035} {arXiv:nucl-th/9803035 [nucl-th]}
  \BibitemShut {NoStop}%
\bibitem [{\citenamefont {Hald}(2000)}]{Hald2000The}%
  \BibitemOpen
  \bibfield  {author} {\bibinfo {author} {\bibfnamefont {A.}~\bibnamefont
  {Hald}},\ }\href@noop {} {\bibfield  {journal} {\bibinfo  {journal}
  {International Statistical Review}\ }\textbf {\bibinfo {volume} {68}},\
  \bibinfo {pages} {137–153} (\bibinfo {year} {2000})}\BibitemShut {NoStop}%
\bibitem [{\citenamefont {Luo}\ and\ \citenamefont {Xu}(2017)}]{Luo:2017faz}%
  \BibitemOpen
  \bibfield  {author} {\bibinfo {author} {\bibfnamefont {X.}~\bibnamefont
  {Luo}}\ and\ \bibinfo {author} {\bibfnamefont {N.}~\bibnamefont {Xu}},\
  }\href@noop {} {\  (\bibinfo {year} {2017})},\ \Eprint
  {http://arxiv.org/abs/1701.02105} {arXiv:1701.02105 [nucl-ex]} \BibitemShut
  {NoStop}%
\bibitem [{\citenamefont {Luo}(2012)}]{Luo:2011tp}%
  \BibitemOpen
  \bibfield  {author} {\bibinfo {author} {\bibfnamefont {X.}~\bibnamefont
  {Luo}},\ }\href {\doibase 10.1088/0954-3899/39/2/025008} {\bibfield
  {journal} {\bibinfo  {journal} {J. Phys.}\ }\textbf {\bibinfo {volume}
  {G39}},\ \bibinfo {pages} {025008} (\bibinfo {year} {2012})},\ \Eprint
  {http://arxiv.org/abs/1109.0593} {arXiv:1109.0593 [physics.data-an]}
  \BibitemShut {NoStop}%
\bibitem [{\citenamefont {Luo}(2015{\natexlab{b}})}]{Luo:2014rea}%
  \BibitemOpen
  \bibfield  {author} {\bibinfo {author} {\bibfnamefont {X.}~\bibnamefont
  {Luo}},\ }\href {\doibase 10.1103/PhysRevC.94.059901,
  10.1103/PhysRevC.91.034907} {\bibfield  {journal} {\bibinfo  {journal} {Phys.
  Rev.}\ }\textbf {\bibinfo {volume} {C91}},\ \bibinfo {pages} {034907}
  (\bibinfo {year} {2015}{\natexlab{b}})},\ \bibinfo {note} {[Erratum: Phys.
  Rev.C94,no.5,059901(2016)]},\ \Eprint {http://arxiv.org/abs/1410.3914}
  {arXiv:1410.3914 [physics.data-an]} \BibitemShut {NoStop}%
\bibitem [{\citenamefont {Alba}\ \emph {et~al.}(2014)\citenamefont {Alba},
  \citenamefont {Alberico}, \citenamefont {Bellwied}, \citenamefont {Bluhm},
  \citenamefont {Mantovani~Sarti}, \citenamefont {Nahrgang},\ and\
  \citenamefont {Ratti}}]{Alba:2014eba}%
  \BibitemOpen
  \bibfield  {author} {\bibinfo {author} {\bibfnamefont {P.}~\bibnamefont
  {Alba}}, \bibinfo {author} {\bibfnamefont {W.}~\bibnamefont {Alberico}},
  \bibinfo {author} {\bibfnamefont {R.}~\bibnamefont {Bellwied}}, \bibinfo
  {author} {\bibfnamefont {M.}~\bibnamefont {Bluhm}}, \bibinfo {author}
  {\bibfnamefont {V.}~\bibnamefont {Mantovani~Sarti}}, \bibinfo {author}
  {\bibfnamefont {M.}~\bibnamefont {Nahrgang}}, \ and\ \bibinfo {author}
  {\bibfnamefont {C.}~\bibnamefont {Ratti}},\ }\href {\doibase
  10.1016/j.physletb.2014.09.052} {\bibfield  {journal} {\bibinfo  {journal}
  {Phys. Lett.}\ }\textbf {\bibinfo {volume} {B738}},\ \bibinfo {pages} {305}
  (\bibinfo {year} {2014})},\ \Eprint {http://arxiv.org/abs/1403.4903}
  {arXiv:1403.4903 [hep-ph]} \BibitemShut {NoStop}%
\bibitem [{\citenamefont {Alba}\ \emph
  {et~al.}(2015{\natexlab{a}})\citenamefont {Alba}, \citenamefont {Bellwied},
  \citenamefont {Bluhm}, \citenamefont {Mantovani~Sarti}, \citenamefont
  {Nahrgang},\ and\ \citenamefont {Ratti}}]{Alba:2015iva}%
  \BibitemOpen
  \bibfield  {author} {\bibinfo {author} {\bibfnamefont {P.}~\bibnamefont
  {Alba}}, \bibinfo {author} {\bibfnamefont {R.}~\bibnamefont {Bellwied}},
  \bibinfo {author} {\bibfnamefont {M.}~\bibnamefont {Bluhm}}, \bibinfo
  {author} {\bibfnamefont {V.}~\bibnamefont {Mantovani~Sarti}}, \bibinfo
  {author} {\bibfnamefont {M.}~\bibnamefont {Nahrgang}}, \ and\ \bibinfo
  {author} {\bibfnamefont {C.}~\bibnamefont {Ratti}},\ }\href {\doibase
  10.1103/PhysRevC.92.064910} {\bibfield  {journal} {\bibinfo  {journal} {Phys.
  Rev.}\ }\textbf {\bibinfo {volume} {C92}},\ \bibinfo {pages} {064910}
  (\bibinfo {year} {2015}{\natexlab{a}})},\ \Eprint
  {http://arxiv.org/abs/1504.03262} {arXiv:1504.03262 [hep-ph]} \BibitemShut
  {NoStop}%
\bibitem [{\citenamefont {Alba}\ \emph
  {et~al.}(2015{\natexlab{b}})\citenamefont {Alba}, \citenamefont {Alberico},
  \citenamefont {Bellwied}, \citenamefont {Bluhm}, \citenamefont {Sarti},
  \citenamefont {Nahrgang},\ and\ \citenamefont {Ratti}}]{Alba:2015lxa}%
  \BibitemOpen
  \bibfield  {author} {\bibinfo {author} {\bibfnamefont {P.}~\bibnamefont
  {Alba}}, \bibinfo {author} {\bibfnamefont {W.}~\bibnamefont {Alberico}},
  \bibinfo {author} {\bibfnamefont {R.}~\bibnamefont {Bellwied}}, \bibinfo
  {author} {\bibfnamefont {M.}~\bibnamefont {Bluhm}}, \bibinfo {author}
  {\bibfnamefont {V.~M.}\ \bibnamefont {Sarti}}, \bibinfo {author}
  {\bibfnamefont {M.}~\bibnamefont {Nahrgang}}, \ and\ \bibinfo {author}
  {\bibfnamefont {C.}~\bibnamefont {Ratti}},\ }\href {\doibase
  10.1088/1742-6596/599/1/012021} {\bibfield  {journal} {\bibinfo  {journal}
  {J. Phys. Conf. Ser.}\ }\textbf {\bibinfo {volume} {599}},\ \bibinfo {pages}
  {012021} (\bibinfo {year} {2015}{\natexlab{b}})}\BibitemShut {NoStop}%
\bibitem [{\citenamefont {Alba}\ \emph {et~al.}(2017)\citenamefont {Alba} \emph
  {et~al.}}]{Alba:2017mqu}%
  \BibitemOpen
  \bibfield  {author} {\bibinfo {author} {\bibfnamefont {P.}~\bibnamefont
  {Alba}} \emph {et~al.},\ }\href@noop {} {\  (\bibinfo {year} {2017})},\
  \Eprint {http://arxiv.org/abs/1702.01113} {arXiv:1702.01113 [hep-lat]}
  \BibitemShut {NoStop}%
\bibitem [{\citenamefont {Noronha-Hostler}\ \emph {et~al.}(2017)\citenamefont
  {Noronha-Hostler}, \citenamefont {Bellwied}, \citenamefont {Gunther},
  \citenamefont {Parotto}, \citenamefont {Pasztor}, \citenamefont {Vazquez},\
  and\ \citenamefont {Ratti}}]{Noronha-Hostler:2016sje}%
  \BibitemOpen
  \bibfield  {author} {\bibinfo {author} {\bibfnamefont {J.}~\bibnamefont
  {Noronha-Hostler}}, \bibinfo {author} {\bibfnamefont {R.}~\bibnamefont
  {Bellwied}}, \bibinfo {author} {\bibfnamefont {J.}~\bibnamefont {Gunther}},
  \bibinfo {author} {\bibfnamefont {P.}~\bibnamefont {Parotto}}, \bibinfo
  {author} {\bibfnamefont {A.}~\bibnamefont {Pasztor}}, \bibinfo {author}
  {\bibfnamefont {I.~P.}\ \bibnamefont {Vazquez}}, \ and\ \bibinfo {author}
  {\bibfnamefont {C.}~\bibnamefont {Ratti}},\ }\href {\doibase
  10.1088/1742-6596/779/1/012050} {\bibfield  {journal} {\bibinfo  {journal}
  {J. Phys. Conf. Ser.}\ }\textbf {\bibinfo {volume} {779}},\ \bibinfo {pages}
  {012050} (\bibinfo {year} {2017})},\ \Eprint
  {http://arxiv.org/abs/1610.00221} {arXiv:1610.00221 [nucl-th]} \BibitemShut
  {NoStop}%
\bibitem [{\citenamefont {Xu}(2017)}]{Xu:2016hxf}%
  \BibitemOpen
  \bibfield  {author} {\bibinfo {author} {\bibfnamefont {J.}~\bibnamefont {Xu}}
  (\bibinfo {collaboration} {STAR}),\ }\href {\doibase
  10.1088/1742-6596/779/1/012073} {\bibfield  {journal} {\bibinfo  {journal}
  {J. Phys. Conf. Ser.}\ }\textbf {\bibinfo {volume} {779}},\ \bibinfo {pages}
  {012073} (\bibinfo {year} {2017})},\ \Eprint
  {http://arxiv.org/abs/1611.07132} {arXiv:1611.07132 [hep-ex]} \BibitemShut
  {NoStop}%
\end{thebibliography}%
\bibliographystyle{apsrev4-1}
\end{document}